# Stochastic equations for thermodynamics

Roumen Tsekov
Department of Physical Chemistry, University of Sofia, 1164 Sofia, Bulgaria

The applicability of stochastic differential equations to thermodynamics is considered and a new form, different from the classical Itô and Stratonovich forms, is introduced. It is shown that the new presentation is more appropriate for the description of thermodynamic fluctuations. The range of validity of the Boltzmann-Einstein principle is also discussed and a generalised alternative is proposed. Both expressions coincide in the small fluctuation limit, providing a normal distribution density.

The controversy concerning the proper meaning and the appropriate application in physics of the stochastic differential equations is full of pitfalls. One is the unclear separation between the macroscopic value of a physical quantity and its fluctuations. The aim of the present paper is to clarify some problems that arise in the application of stochastic differential equations in thermodynamics. It is well-known [1] that many processes in nature can be described by the following stochastic equation

$$dX = A(X)dt + B(X)dW(t)$$

named after Paul Langevin. Here $X$ is the quantity under observation, $t$ is time, $A$ and $B$ are deterministic functions and $W$ is a random Wiener process. The integral form of the Langevin equation is

$$X(t) = X(0) + \int_0^t A(X)ds + \int_0^t B(X) ]\alpha[ \, dW \qquad (1)$$

where the notation $]\alpha[$ indicates a peculiar definition of the integrals over a Wiener process.

As was mentioned, the main problem in eq. (1) is the interpretation of the last integral. Following standard mathematics, it can be expressed by a Riemann sum as

$$\lim_{N \to \infty} \sum_{n=0}^{N-1} B[X(\frac{n+\alpha}{N}t)][W(\frac{n+1}{N}t) - W(\frac{n}{N}t)]$$

where $\alpha$ is a real number between 0 and 1. In contrast to the usual integrals, which are independent of the value of $\alpha$, this expression depends substantially on the choice of the middle point in the time intervals. This is evident from the corresponding Fokker-Planck equation describing the evolution of the probability density $P(x,t)$

$$\partial_t P = \partial_x[-AP + (B^{2\alpha}/2)\partial_x(B^{2(1-\alpha)}P)] \tag{2}$$

which can be rigorously derived from eq. (1) [1]. In the literature, there are two world-wide accepted choices for $\alpha$ leading to two different forms of eq. (2): the Itô form with $\alpha = 0$ [2] and the Stratonovich form with $\alpha = 1/2$ [3]. From the mathematical point of view, both forms are correct but their application to physics and chemistry generates some problems [4]. The main difficulty is related to the right attribution of physical meaning to the functions $A$ and $B$ [5]. The latter are not independent and their relationship, being the subject of the fluctuation-dissipation relation [6], is given by the equilibrium solution $\bar{P}(x)$ of eq. (2)

$$2A = (1-\alpha)\partial_x B^2 + B^2 \partial_x \ln \bar{P}$$

As seen, the $\alpha$-problem disappears if $B$ is constant. For this reason, let us start with a pure random process obeying the following Langevin equation

$$dX = A(X)dt + dW \tag{3}$$

Regardless the value of $\alpha$, the corresponding Fokker-Planck equation is $\partial_t P = \partial_x(-AP + \partial_x P/2)$ and the drift term $A$ is related to the equilibrium probability density as follows

$$2A = \partial_x \ln \bar{P} \tag{4}$$

Let us now introduce a new random variable $Y$ being a deterministic function of $X$. The equilibrium probability density $\bar{P}(x)$ is connected to $\bar{P}(y)$ via the relation [7]

$$\bar{P}(x) = \bar{P}(y)C(y) \tag{5}$$

where $C(Y) = d_X Y$. Multiplying eq. (3) by $C$ and using eqs. (4) and (5) one obtains a new stochastic differential equation for the random evolution of $Y$ driven by a multiplicative white noise

$$dY = (C^2/2)(d_Y \ln \bar{P}C)dt + C]_\alpha[ dW \tag{6}$$

Equation (6) is a particular sample of the Langevin equation with $B = C$. Its corresponding Fokker-Planck equation is

$$\partial_t P = \partial_y [-C^2 P d_y \ln \bar{P}C + C^{2\alpha}\partial_y (C^{2(1-\alpha)} P)]/2$$

which is correct in the equilibrium state $(P = \bar{P})$ only if $\alpha = 1/2$. Hence,

$$\partial_t P = \partial_y (C^2 P \partial_y \ln P/\bar{P})/2 \tag{7}$$

and the correct form of eq. (6) is the Stratonovich one

$$dY = (C^2/2)(d_Y \ln \bar{P}C)dt + C\,]1/2[\,dW \tag{8}$$

The fact that the Stratonovich form corresponds to the usual rules of mathematics is proven by the Wong-Zakai theorem [8]. However, there are many alternative forms of eq. (6) which provide the same Fokker-Planck equation. Two examples are

$$dY = (C^2/2)(d_Y \ln \bar{P}C^2)dt + C\,]0[\,dW \tag{9}$$

$$dY = (C^2/2)(d_Y \ln \bar{P})dt + C\,]1[\,dW \tag{10}$$

Equation (9) is the Itô form of the Langevin equation. Its advantage is that the equilibrium average value of the drift term is zero. For this reason it is suitable for juxtaposition with the corresponding macroscopic equations [7]. Equation (10) is a new one. Its convenience is related to the proportionality of the drift term to the gradient of the equilibrium distribution density. In this sense, it could be appropriate for non-equilibrium thermodynamics. All equations (8-10) are exact. There is no physics there and using any one of them is a matter of convenience. In general, the transition to another $\alpha$, without change in the corresponding Fokker-Planck equation, is ruled by $\partial_\alpha (B)\alpha[\,dW) = (dB/dX)Bdt$ [1].

According to thermodynamics, the characteristic function of a closed system at constant temperature $T$ is the free energy $F$. Any spontaneous process in the system leads to decrease of $F$ to its minimal value corresponding to the equilibrium state. Hence, if $Z$ is a variable parameter of the system, the rate of free energy decrease can be presented as

$$d_t F = (\partial_Z F) d_t Z \leq 0 \tag{11}$$

On the other hand, according to the non-equilibrium thermodynamics, there is a linear relationship between the rate change of $Z$ and the gradient of the free energy

$$d_t Z = -M(Z)\partial_Z F \qquad (12)$$

where $M^{-1}$ is the resistance of the system. Owing to the positive definition of $M$ inequality (11) is always fulfilled. Equation (12) is not stochastic and describes only the irreversible evolution toward equilibrium without accounting for the fluctuations. The latter are not subject of the second law of thermodynamics.

The difference between the equilibrium free energy and the conditional free energy for a given value of $Z$ is proportional to the logarithm of the equilibrium probability density to observe this fluctuation value [6]

$$F(\bar{Z}) - F(Z) \propto k_B T \ln \bar{P}(Z) \qquad (13)$$

Introducing this expression in eq. (12) one can easily obtain that the differential stochastic equation for the $Z$-fluctuations should be written in the form

$$dZ = k_B TM(d_Z \ln \bar{P})dt + \sqrt{2k_B TM}\, ]1[\, dW \qquad (14)$$

Equation (14) is a particular sample of eq. (10) and the corresponding Fokker-Planck equation is

$$\partial_t P = \partial_z (k_B TMP \partial_z \ln P/\bar{P})$$

As is seen, the most appropriate value of $\alpha$ in the description of thermodynamic fluctuations is $\alpha = 1$ and this is not surprising. In physics, the casualty principle is very important. The only value of $\alpha$ which accounts for the whole influence of the Wiener process to the contemporary evolution of the variable under observation is 1. Other $\alpha$-values correspond to a physically unacceptable influence of fluctuations taking place after the current time to the moment state of the considered process. A common mistake here is to treat eq. (12) as the average product of eq. (14). This is only true if $M$ is constant. Equation (12) is not exact. It is a result of the Second Law and does not take into account the thermodynamic fluctuations which lead to the free energy increase. The exact equation is (14) which reduces to eq. (12) in the absence of fluctuations.

Finally, we shell pay attention to a problem of range of validity of the Boltzmann-Einstein principle (13). It is obvious that the Boltzmann-Einstein principle is not general because it is not invariant against non-linear change of the variable by the law (5). The only case for which it is always satisfied is the small fluctuation limit corresponding to a Gaussian distribution function.

Hereafter, an alternative point of view is presented which seems to be more general. According to the statistical mechanics, the equilibrium probability density for fluctuations of $Z(\Omega)$ can be calculated by the canonical Gibbs distribution

$$\bar{P}(z) = \int \delta(z-Z) \exp(\frac{F-H}{k_B T}) d\Omega$$

where $H(\Omega)$ is Hamilton function of the system and the free energy $F$ is given by

$$\exp(-\frac{F}{k_B T}) = \int \exp(-\frac{H}{k_B T}) d\Omega \qquad (15)$$

A more appropriate statistical quantity for description of the fluctuations is the characteristic function $\Theta$ which is defined as the Fourier image of the probability density

$$\Theta(s) \equiv \int \exp(isz) \bar{P}(z) dz = \exp(\frac{F}{k_B T}) \int \exp(-\frac{H - isk_B TZ}{k_B T}) d\Omega \qquad (16)$$

According to the thermodynamics, the average value $\bar{Z}$ of the fluctuating thermodynamic parameter is equal to the first derivative of the free energy $F$ with respect to its thermodynamically conjugated quantity $\zeta$, i.e. $\bar{Z} = \partial_\zeta F$. For instance, in simple systems $\zeta$ could be temperature, volume or number of particles and the corresponding $Z$ quantities are entropy, pressure or chemical potentials. From this point of view and eq. (15), the characteristic function (16) can be expressed as

$$\Theta(s) = \exp[\frac{F(\zeta) - F(\zeta - isk_B T)}{k_B T}]$$

and an alternative to the Boltzmann-Einstein principle (13) is $F(\zeta) - F(\zeta - isk_B T) = k_B T \ln \Theta(s)$ which is invariant to any change of the variable. It is clear that for Gaussian fluctuations both the expressions coincide. Owing to the minimal value of the free energy at equilibrium, the necessary condition $0 \leq \Theta \leq 1$ follows from the equation above. The logarithm of the characteristic function is the so-called cumulant generating function [6] using which a number of correlation characteristics of the system fluctuations can be calculated.

The results obtained in the present paper possess rather methodological character rather than being a concrete new knowledge for Nature. The foregoing discussion aims to acquaint the readers with the common difficulty arising in the use of stochastic differential equations and to prevent them of undesirable mistakes. Since the theory of the equilibrium state is much more developed, the basic checkpoint for any kinetic theory is the reproduction of the equilibrium theory results. The reader can note that this is a good tool for discrimination between the different interpretations of the stochastic equations. Of course, the latter can be obtained from first principles. However, owing to the complexity of the systems this is possible only in very simple cases, e.g. harmonic oscillators.

In a previous paper [9] starting from the classical mechanics, we have demonstrated that the dynamics of a Brownian particle in solids obeys the following equation

$$dR = -M(R)(\partial_R U)dt + \sqrt{2k_B TM(R)}\,]1[\,dW$$

where $R$ is the particle coordinate, $U(R)$ is potential and $M(R)$ is position dependent mobility. As is seen, this equation is a particular example of eq. (14). Any other choice of $\alpha$ will reflect in physically undue dependence of the equilibrium distribution on the particle friction.

## Appendix

Nowadays, the choice $\alpha = 1$ is known as the Hänggi-Klimontovich form [10]. The goal of the present appendix is to apply the above theory to another interesting example, which is related to dissipative nonlinearity. At large velocities the linear friction regime is usually violated [11] and the friction coefficient of a particle becomes velocity-dependent. A very generic model follows from the theory of activated transport, where the velocity of particle migration under the action of a force $f$ is given by

$$v/c = \exp(f\lambda/k_B T)/2 - \exp(-f\lambda/k_B T)/2 = \sinh(f\lambda/k_B T) \qquad (17)$$

Here $c$ is the average thermal velocity and $\lambda$ is the typical length scale of the environment structure or the mean free path in gases. Reversing eq. (17) yields an expression for the friction force

$$f = (mc^2/\lambda)\operatorname{arcsinh}(v/c) \qquad (18)$$

It is interesting that the friction coefficient $\gamma \equiv f/mv$ decreases with increase of the particle velocity. At slow motion $v \ll c$ it is nearly constant $\gamma \approx c/\lambda$, while at large velocities the friction force is logarithmically dependent on $v$, i.e. the Coulomb-Amontons law of solid friction almost holds. In the middle range of velocities the friction force is well approximated by the cubic friction model with $\gamma = (c - v^2/6c)/\lambda$. Equation (17) is an example of a nonlinear flow-force relationship from non-equilibrium thermodynamics. Accordingly, the stochastic differential equation of a free Brownian particle with nonlinear friction coefficient acquires the form

$$dV = -\gamma(V)Vdt + B(V)]\alpha[\,dW \tag{19}$$

As is seen, the thermal noise in eq. (19) is multiplicative and the corresponding Fokker-Planck equation reads

$$\partial_t P = \partial_v[\gamma(v)vP + (B^{2\alpha}/2)\partial_v(B^{2(1-\alpha)}P)] \tag{20}$$

The equilibrium solution of the Fokker-Planck equation should be the Maxwell distribution. Substituting $\bar{P}(v) = \exp(-v^2/2c^2)/\sqrt{2\pi}c$ in eq. (20) provides the fluctuation-dissipation relation

$$B^2 - 2(1-\alpha)c^2 \partial_{v^2} B^2 = 2c^2\gamma(v) \tag{21}$$

Thus, the diffusion coefficient in the velocity space $B^2$ depends essentially on the choice of the middle point. In the Itô case $\alpha = 0$ the integration of eq. (21) yields

$$B_0^2 = \exp(v^2/2c^2)\int_{v^2}^{\infty}\gamma(v)\exp(-v^2/2c^2)dv^2 = 2\bar{P}^{-1}\int_v^{\infty}\gamma(v)v\bar{P}dv \tag{22}$$

Therefore, the Itô diffusion coefficient equals to the friction force averaged for larger velocities over the Maxwell distribution. On the contrary, the Hänggi-Klimontovich diffusion coefficient is purely local, since eq. (21) reduces to $B_1^2 = 2c^2\gamma(v) = 2k_B T\gamma(v)/m$ at $\alpha = 1$. Consequently, the choice of the middle point is physically evident, which discriminates the stochastic models. The effect of nonlinear friction on the quantum Brownian motion was recently described [12].